\documentclass[twocolumn,showpacs,prl]{revtex4}

\usepackage{graphicx}%
\usepackage{dcolumn}
\usepackage{amsmath}
\usepackage{latexsym}

\begin {document}

\title
{
Directed Fixed Energy Sandpile Model
}
\author
{
R. Karmakar and S. S. Manna
}
\affiliation
{
Satyendra Nath Bose National Centre for Basic Sciences 
    Block-JD, Sector-III, Salt Lake, Kolkata-700098, India
}

\begin{abstract}
   We numerically study the directed version of the fixed energy sandpile.
On a closed square lattice, the dynamical evolution of a fixed density of sand grains is
studied. The activity of the system shows a continuous phase transition around
a critical density. While the deterministic version has the set of nontrivial exponents, the
stochastic model is characterized by mean field like exponents.
\end{abstract}
\maketitle

    Spontaneous emergence of long ranged spatio-temporal correlations 
    under a self-organizing dynamics, in absence of a fine tuning parameter 
    is the basic idea of Self-organized Criticality (SOC).
    Sandpile models are the prototype models of SOC
    \cite {Bak,Bakbook,Dhar1,Grass,Maya, Prie,Mohanty,Stella,Sree}.
    In sandpile 
    models an integer height variable $h_i$ representing the number of grains in the sand column is associated 
    with every site of a regular lattice. 
    The system is driven by adding unit grains of sand
    at a time. When the height $h_i > h_c$ the sand column
    topples and it looses some grains which are distributed among
    the neighboring sites \cite {Bak}. This creates an avalanche of sand column topplings 
    and the extent of such activity measures the size of the avalanche.
    Sand grains go out of the system through a boundary so that in the steady state the fluxes of
    inflow and outflow currents balance.

       On the other hand a fixed energy sandpile (FES) \cite {Rossi, Munoz, Vespignani, Dickman} is a
    sandpile model within a closed system. 
    Therefore the total mass of sand in this system is a conserved quantity.
    The control parameter is the density $\zeta$ of grains.
    A stable state has heights varying from 0 to $h_c-1$
    and is called an $inactive$ state, whereas, any height configuration 
    that has at least one unstable site is said to be in the
    $active$ state. The dynamics of the system starts with a random distribution of 
    $N=\zeta L^2$ grains. Initially some sites may be unstable, which topple. 
    Consequently some of the neighboring sites may topple again and the activity continues. 
    For an infinitely large system there exists a critical threshold $\zeta_c$ such that 
    if $\zeta < \zeta_c$ the activity terminates and the system gets absorbed in an 
    inactive state where as for $\zeta > \zeta_c$ the activity of the system fluctuates 
    but maintains a steady mean value \cite {Rossi}. Therefore
    $\zeta_c$ is the critical point 
    of a continuous phase transition from an absorbed phase to an active phase.

        After the dynamics starts, the system takes some time to relax to the steady 
   state. The activity at a certain time is measured by the fraction $\rho$ of
   lattice sites which are unstable at that time. In general the mean activity $\langle \rho \rangle$
   is a function of the density $\zeta$ i.e., the deviation from the critical point
   $\Delta = \zeta - \zeta_c$ and also the system size $L$. The simultaneous dependence
   of activity on $\Delta$ and $L$ is expressed by the following scaling form:
\begin {equation}
\langle \rho (\Delta,L) \rangle = L^{-\beta/\nu_{\perp}}{\cal G}(L^{1/\nu_{\perp}}\Delta)
\end {equation}
   where ${\cal G}(x)$ is an universal scaling function such that ${\cal G}(x)\rightarrow x^{\beta}$
   when $x >>1$. This implies that for a certain range of $\Delta$ if $L$ is so large
   that $L^{1/\nu_{\perp}}\Delta >>1$ then $\langle \rho (\Delta,L) \rangle$ is independent of
   $L$ and depends solely on $\Delta$ as $\langle \rho (\Delta,L) \rangle \sim \Delta^{\beta}$.
   Naturally $\beta$ is the order parameter exponent for the transition. On the other
   hand when $x<<1$, ${\cal G}(x) \rightarrow $ constant, independent of both $\Delta$ and $L$
   implies that right at the critical point $\zeta = \zeta_c$ the order parameter varies with
   the system size as: $\langle \rho (L) \rangle \sim L^{-\beta/\nu_{\perp}}$, independent of $\Delta$.

      To which universality class the FES model exponents should correspond to?  
   Intensive research has been done to study the universality class of the phase
   transitions in FES models.
   It has been suggested that FES belongs to the universality
   class of the linear interface models (LIM) but not of that of the Directed percolation
   (DP) universality class \cite {Rossi, Munoz, Vespignani, Dickman}. 
   DP is generic for continuous absorbing state transitions in the absence of a conservation law
   where as in FES there exists a conserved field which is the density and
   it couples the order parameter, i.e., the mean activity.
   We, in the present study like to examine
   if an explicit application of the directional bias to the FES system makes the
   system behave as DP or it results to another new universality class.
   
      Application of a global directional bias onto a system has been proved to have
   strong effect on the critical behaviors of various models in Statistical Physics.
   In directed systems, degrees of freedom of the individual elements is reduced,
   which shrinks the configuration space of the system
   compared to undirected system. As a result a directed system is simpler
   and easily tractable analytically.  Examples include
   Directed percolation \cite {DP},
   Directed Sandpile Model \cite {DSM, Pastor,vesp1}, Directed River networks \cite {DRN}
   and Directed Self-avoiding walks \cite {DSAW} etc.

       We study here the directed fixed energy sandpile (DFES) models on an 
    oriented square lattice placed on the $x-y$ plane, periodic boundary conditions 
    are imposed along both the directions. A preferred direction is imposed onto this system
    along the $-y$ direction. The critical height of stability of a sand 
    column is fixed at $h_c=1$, and on a toppling two grains of sand are distributed
    to the two neighboring sites along the preferred direction i.e., at the
    lower-left (LL) and lower-right (LR) positions.

       Two different versions of the model are studied according to the
    rule by which the two grains are distributed in a toppling: 
    (i) both the LL and LR sites get one grain each, which we call as the 
    deterministic directed fixed energy sandpile model (DDFES),
    (ii) each of the two grains is distributed randomly to any of the LL 
    or LR positions, and this version is called the stochastic directed fixed energy 
    sandpile (SDFES) model. 

\begin{figure}[top]
\begin{center}
\includegraphics[width=5.5cm]{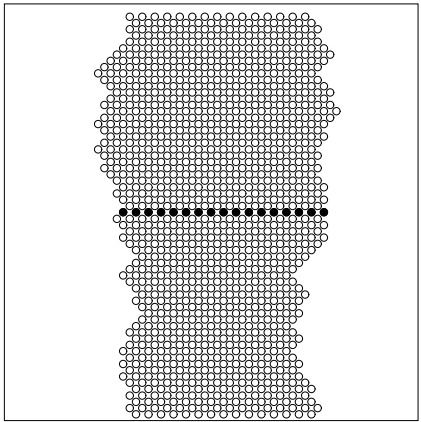}
\end{center}
\caption{
Filled circles denote the sites on the toppling front in an infinite avalanche of
DDFES on an oriented square lattice of size $L=32$. This avalanche is periodic
and has a period 64. Horizontal empty circles denote positions of the TF in another
63 time units. The two end sites of the TF fluctuates but maintains a mean distance 
of $L/2$.
}
\end{figure}

       We shall first discuss the properties of DDFES. The critical point
    can be arrived at from an inactive state by adding grains one by one on an
    initial empty lattice followed by the relaxation of the avalanche.
    On the average both the size and the life times of the avalanches increase as the density grows.
    The toppling front (TF) of an avalanche is a set of horizontal contiguous sites
    which travels downward with unit speed. The length of the TF however fluctuates (Fig. 1).
    If at an intermediate time the TF has $n$ sites, then at the next time step
    its length can be only $n-1, n$ or $n+1$. For a finite avalanche the TF
    first grows from a single site to a certain length and then shrinks to
    zero. The set of sites covered by the left and right end sites of TF are 
    the paths of two annihilating random walkers \cite {DSM}. 
    For a finite avalanche they meet and annihilate, however for an
    infinite avalanche these two random walkers cannot meet and the best possible
    way it can be ensured if they can maintain a distance of $L/2$ on the average from each other.

        Therefore the minimum possible sustained activity for DDFES in a system
    of size $L$ is $\rho=1/(2L)$. What is interesting is, on such a system if the density is slightly
    increased the avalanches created by the additional grains die away and the
    system maintains the activity of the infinite avalanche. In a sense the system
    gets locked with this activity for a certain range of grain density. However on 
    increasing the density even further, a second infinite avalanche is created
    and both the TFs run simultaneously resulting a sudden jump in the
    activity by doubling its magnitude to $1/L$. This continues for some range of
    grain density which ends at another jump in activity to $3/(2L)$. Thus
    in general the variation of activity is discrete and has a step like
    variation with step heights $1/(2L)$ (Fig. 2). As the system size increases the step
    height decreases to zero and the variation of $\rho$ with $\zeta$
    becomes more and more smooth. Similar step-like behavior
    of the order parameter was also observed in \cite {bagnoli}.

      The configurations at the steady state are periodic and
    the same detailed distribution of grain numbers at all sites
    repeat at regular intervals of time. The periodic time is always multiples
    of $L$ in a $L \times L$ system. For small systems this period has 
    different values for different initial configurations but in most (about
    95\%) cases the period is $2L$ and rarely $L$, $3L$, $4L$ etc. However,
    for bigger system sizes e.g., for $L$ = 512, 1024 or more the period is
    always $2L$. This helps to calculate the order parameter. Given an initial 
    distribution of grains, it therefore needs to find out the mean activity 
    over only a period and then average over many initial configurations.

\begin{figure}[top]
\begin{center}
\includegraphics[width=6.5cm]{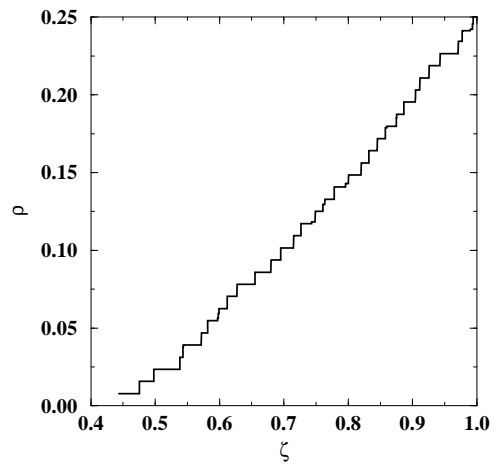}
\end{center}
\caption{
The mean activity $\langle \rho \rangle$ in a system of size $L=64$ 
which grows with the density in a step like manner where the step heights
are $\Delta \rho = 1/(2L)$.
}
\end{figure}

      The critical density $\zeta_c$ actually has a system size dependence.
   To study this variation we start with a closed empty lattice of size $L \times L$ and go on adding
   sand grains one by one at randomly selected lattice sites similar to what is done in an open sandpile. The dynamics
   of the avalanche is followed for each sand grain added. The mean avalanche size increases
   with the density of grains in the system and at a certain $\zeta=\zeta_c(L)$ depending on the
   sequence of randomly selected sites at which the grains were dropped, the activity does not stop
   any more and an ``infinite'' avalanche continues for ever. In practice, in our simulation we
   followed an avalanche up to a certain relaxation time $T=10^6$ for $L < 128$ and $T= 5 \times 10^6$ 
   for $L \ge 128$ to declare the avalanche as infinite. Repeating this simulation a large number of
   times, every time starting from an empty system, we calculate the average critical density
   $\langle \zeta_c(L) \rangle$. These values are then extrapolated as: 
   $\langle \zeta_c(L) \rangle =\zeta_c+AL^{-1/\nu_{\parallel}}$ as shown in Fig. 3 to obtain 
   $\zeta_c = 0.4115 \pm 0.002$ and $1/\nu_{\parallel}=0.361$ giving $\nu_{\parallel} \approx 2.77$. 
   We also esimated $\zeta_c$ by the scaling plot of mean avalanche size $\langle s(L) \rangle L^{-0.05}$
   vs. $\Delta L^{0.03}$ which is also consistent with our estimate of $\zeta_c = 0.4115$.

\begin{figure}[top]
\begin{center}
\includegraphics[width=6.5cm]{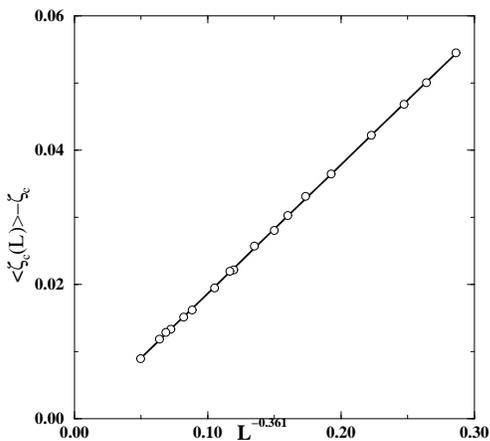}
\end{center}
\caption{
The variation of the deviation of the critical density $\zeta_c(L)$ of a system of size $L$ 
from its value $\zeta_c$ at the infinitely large system 
is plotted with $L^{-1/\nu_{\parallel}}$. For DDFES we obtain 
$\zeta_c \approx 0.4115$,$\nu_{\parallel}=1/0.361 \approx 2.77$.
}
\end{figure}

      A system of size $L$ is filled initially with density $\zeta$ by randomly distributing $\zeta L^2$ grains. 
   We allow the system to evolve up to a relaxation time $T$ after which the activity is measured
   at every time step. To measure the mean activity $\langle \rho(\Delta, L) \rangle$ for a slightly
   higher density $\zeta + \delta \zeta$ we take the advantage of the fact that the system dynamics is deterministic.
   On the same initial distribution of grains corresponding to the density $\zeta$ another
   $(\delta \zeta) L^2$ grains are randomly added. This ensures that if certain density gives sustained activity,
   its higher density necessarily gives a non-stop activity. These measurements are then repeated for
   different system sizes. In Fig. 4 we show the scaling of the order parameter
   on double logarithmic scale. Plotting $\langle \rho(\Delta, L) \rangle L$ with
   $L^{0.55} \Delta$ we observe a nice data collapse for system sizes $L$ =128, 256 and 512.
   Comparing with the Eqn. 1 we conclude $\beta/\nu_{\perp}$ = 1 and $1/\nu_{\perp}$=0.55.
   This implies that $\beta = \nu_{\perp} \approx 1.82$. 

      The analysis so far enables us to estimate the dynamical exponent $z = \nu_{\parallel}/\nu_{\perp}
   \approx 1.52$. This value of the dynamical exponent is directly verified by measuring the
   survival probability. The survival probability $P(t)$ that the initial activity in a random 
   distribution of grains survives a time $t$ has an exponential distribution as: 
   $P(t) \sim exp(-t/\tau)$. At the critical point $\zeta_c$
   the characteristic time is a function of only the system size as: $\tau(L) \sim L^z$
   where, $z$ is the dynamical exponent $z=\nu_{\parallel}/\nu_{\perp}$ of the system. Therefore we
   calculate the average survival time $\langle t_c(L) \rangle$ which is also proportional to 
   $L^z$ at $\zeta_c$ for different system sizes. This is done again by dropping grains of sand one by one
   into a closed system and calculating the life time of the largest avalanche
   before the system gets locked into an infinite avalanche. Averaging over many initial configurations
   the largest life-time $\langle t_c(L) \rangle$ is plotted
   in Fig. 5 for on a double logarithmic scale. The slope of the straight line 
   gives the value for the dynamical exponent $z=1.49 \pm 0.05$ for DDFES compared to 1.52
   obtained previously.

\begin{figure}[top]
\begin{center}
\includegraphics[width=6.5cm]{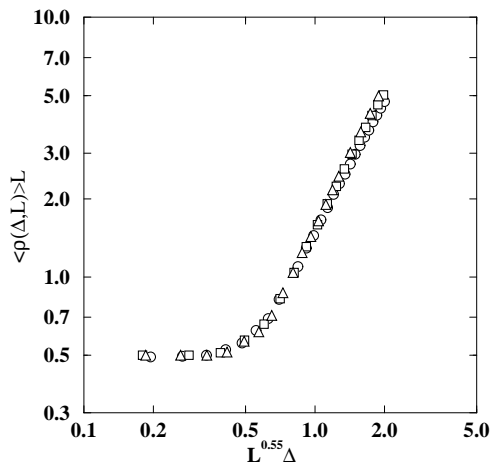}
\end{center}
\caption{
Scaling of the order parameter $\langle \rho(\Delta,L)\rangle$ with the
deviation $\Delta = \zeta-\zeta_c$ from the critical point. From this data
collapse and Eqn. (1) we find, $\beta/\nu_{\perp}=1$ and $1/\nu_{\perp}=0.55$.
}
\end{figure}

      In the stochastic directed fixed energy sandpile model
   the critical density $\zeta_c$ is found to be very close to 0.5. The
   order parameter has a highly linear variation with $\Delta$ as: $\rho(\Delta) = A\Delta$ where
   $A \approx 0.46$. A similar to DDFES calculation of the system size dependent critical density
   after extrapolation $\langle \zeta_c(L) \rangle =\zeta_c+A'L^{-1/\nu_{\parallel}}$
   gives $\nu_{\parallel} \approx 1$ and $A' \approx 2.4$. The plot of 
   $\langle \rho(\Delta,L) \rangle $ vs. $\Delta$ is a very nice straight line and
   the plot of data for different system sizes fall on top of one another. We conclude
   that $\beta \approx 1$ and $\nu_{\perp} \approx \infty$.

\begin{figure}[top]
\begin{center}
\includegraphics[width=6.5cm]{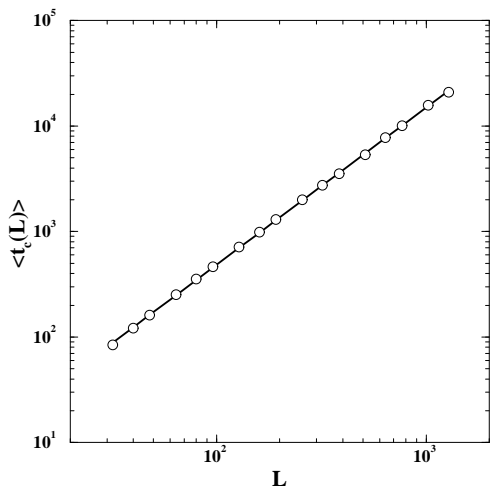}
\end{center}
\caption{
The average of maximal life-time $\langle t_c(L) \rangle$ of the avalanche prior to the infinite avalanche
is plotted with $L$. The slope gives a measure of the dynamical exponent $z=1.49$.
}
\end{figure}

      The roughneing of the associated interface in our FES models is studied.
   If $H_i(L,t)$ denotes the number of topplings upto the time $t$ then the set of $H_i(L,t)$s for all $i$
   represents an interface. For DDFES the width of this interface fluctuates periodically
   but its average grows
   as: ${W(L,t)} \sim L^{\alpha}{\cal G}(t/L^z_i)$ where we find
   for $\alpha=0.31$ and $z=1.6$ in comparison to the Linear Interface Model results
   $\alpha=0.75$ and $z=1.56$ \cite {LIM}. 

      Finally we study the DFES models on oriented square lattices of rectangular shapes,
   the longer sides being parallel to the preferred direction. For DDFES model the TFs are 
   contiguous sites covering the transverse direction in the form of rings. These toppling rings
   are perfectly stable, once formed they never change in shape. As density increases, the
   number of such rings increases. On the other hand for SDFES, the toppling sites are
   randomly scattered throughout the system (Fig. 6).

      To summarize, we have studied the directed version of the fixed energy sandpile on
   the oriented square lattices. Like isotropic FES, our directed FES also shows a continuous
   phase transition from an absorbed phase to an active phase. Two versions of the model are studied. 
   In the deterministic FES, the grain number configurations are periodic and repeats at regular
   time interval of $2L$. For this model the critical points as well as the critical exponents 
   are found to be non-trivial and belong to a new universality class. The other version has 
   the stochastic toppling dynamical rules and exponents of mean-field nature are found for 
   this model.

\begin {center}
\begin {table}
\begin {tabular}{lllll} \\ \hline \hline
Model       &  $\zeta_c$   & $\beta$  & $\nu_{\perp}$ & $\nu_{\parallel}$    \\ \hline
DDFES       &  0.4115      & 1.82     & 1.82          &  2.77              \\ 
SDFES       &  0.5         & 1.00     & $\infty$      &  1.00                   \\
BTW FES     &  2.125       & 0.7      &  0.90         &  1.49              \\
Manna FES \hspace*{0.3cm}   &  0.71695 \hspace*{0.3cm}    & 0.64     &  0.82         &  1.29              \\
DP          &              & 0.583    & 0.733         & 1.295   \\ \hline
\end {tabular}
\caption
{
Comparison of critical points and exponents for different models of fixed energy sandpiles.
Exponent values for BTW and Manna sandpiles are taken from \cite {Vespignani}, DP exponents
from \cite {DP1}.
}
\end{table}
\end {center}

\begin{figure}[top]
\begin{center}
\includegraphics[width=6.5cm]{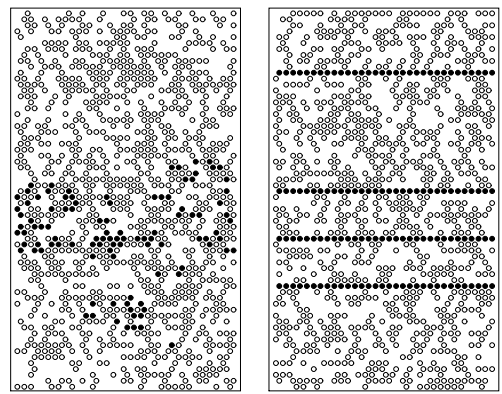}
\end{center}
\caption
{
Two snapshots of height configurations in the directed fixed energy sandpile
model on a $32 \times 64$ oriented square lattice, downward direction being the
preferred direction. Stochastic DFES is shown on the left where as the deterministic
DFES is shown on the right. Active sites are shown by filled circles, open circles
denote sites with height 1 and vacant sites are not indicated.
}
\end{figure}
      
   We thank F. O. Ogundare for some initial discussions.

\end{document}